# Quantum states tomography with noisy measurement channels[1]

Yu.I. Bogdanov[2a,b,c], B.I. Bantysh[a,b], N.A. Bogdanova[a,b], A. B. Kvasnyy[a,c], V.F. Lukichev[a]

[a]Institute of Physics and Technology, Russian Academy of Sciences;
[b]National Research University of Electronic Technology MIET;
[c]National Research Nuclear University MEPHI

**ABSTRACT**

We consider realistic measurement systems, where measurements are accompanied by decoherence processes. The aim of this work is the construction of methods and algorithms for precise quantum measurements with fidelity close to the fundamental limit. In the present work the notions of ideal and non-ideal quantum measurements are strictly formalized. It is shown that non-ideal quantum measurements could be represented as a mixture of ideal measurements. Based on root approach the quantum state reconstruction method is developed. Informational accuracy theory of non-ideal quantum measurements is proposed. The monitoring of the amount of information about the quantum state parameters is examined, including the analysis of the information degradation under the noise influence. The study of achievable fidelity in non-ideal quantum measurements is performed. The results of simulation of fidelity characteristics of a wide class of quantum protocols based on polyhedrons geometry with high level of symmetry are presented. The impact of different decoherence mechanisms, including qubit amplitude and phase relaxation, bit-flip and phase-flip, is considered.

## 1. QUANTUM STATES TOMOGRAPHY

An important element of quantum information technology realization is the quality control for quantum states tomography. Herewith, we consider tomography to be the procedure, consisting of two stages. The first one is the collection of statistical data about the quantum state parameters. This stage is based on Niels Bohr's complementarity principle, which means that one needs to perform a set of complementary measurements to obtain the complete information about the quantum state [1, 2]. The quantum measurements theory is based on von Neumann measurements [3] and their generalizations [4]. On the second stage of tomography the mathematical procedure of quantum state reconstruction, using the measurements results, is performed. The methodology of reconstruction, that we are using in this work, is based on the maximum likelihood estimation proposed by Ronald Fisher [5-10]. The root approach to estimation of quantum state parameters uses this method and has optimal asymptotic properties [11-13].

Any real quantum measurement system is open. Here, along with the interaction of the quantum state and measurement set-up, there exists an interaction between quantum state and its environment. This results in the exchange of information between them and the quantum state decoherence, which leads to a decrease in reconstruction fidelity. The description of decoherence processes could be presented by means of reduced dynamics of open quantum systems, which is based on the concept of complete positivity [14-18]. In the current work we study the impact of decoherence on the results of quantum states measurement.

The quantum tomography protocol is described with the so-called instrumental matrix $X$ [11-13]. In the simplest case, each row $X_j$ of this matrix sets the bra-vector in the Hilbert space. The state under study is projected on these vectors. Let the state under study be described with the density matrix $\rho$. The probability of the registration of $j$-th event is set by the following statement:

$$\lambda_j = X_j \rho X_j^+ = \mathrm{Tr}\left(X_j^+ X_j \rho\right) = \mathrm{Tr}\left(\Lambda_j \rho\right), \quad j = 1, \ldots, m, \qquad (1)$$

---

where $\Lambda_j = X_j^+ X_j$ are the measurement operators (projectors), $m$ is the number of protocol rows (the total amount of measuring projections). In the case of a pure input state, which is described with vector state $c$, the statement (1) could be rewritten in another form: $\lambda_j = c^+ \Lambda_j c$.

Each row of instrumental matrix is included in the protocol with its weights. Let $t_j$ be the weight (time of measurement) of $j$-th row. Let us normalize the total weighted probability to the sample size ($n$ is the total sample size by all protocol rows):

$$\sum_{j=1}^{m} t_j \lambda_j = n. \qquad (2)$$

In the present work we limit ourselves to the protocols which form the decomposition of unity. In this case the sum of all $t_j \Lambda_j$ matrices is proportional to the identity matrix:

$$\sum_{j=1}^{m} t_j \Lambda_j = nI. \qquad (3)$$

Here $I$ is the identity matrix. Note that condition (3) together with vector-state normalization condition $\langle c|c \rangle = 1$ directly leads to (2).

In real experimental conditions it is important to get a high fidelity of the quantum state reconstruction, having only a limited sample of size $n$ from the quantum statistical ensemble representatives. To achieve this goal one can use the procedures of quantum state reconstruction using the root approach and maximum likelihood estimation proposed in our works [11-13].

Let us call measurements, which can be described with (1), as "pure" (or "clear"). Formally, in the more general case each matrix $\Lambda_j$ is a some positive-definite operator that could describe more complicated measurements which cannot be reduced to projective measurements:

$$\Lambda_j^{mixed} = \sum_q f_j^q X_j^{q+} X_j^q, \quad f_j^q > 0, \quad \sum_q f_j^q = 1. \qquad (4)$$

Measurements, which can be described with these operators, will be herewith called "mixed" (or "fuzzy"). In these measurements with probabilities $f_j^q$ different projections measurements are realized. Indeed, substitution of operator $\Lambda_j^{mixed}$ to (1) gives us the law of total probability for the realization of event $j$:

$$\lambda_j = \text{Tr}\left(\Lambda_j^{mixed} \rho\right) = \sum_q f_j^q \text{Tr}\left(X_j^{q+} X_j^q \rho\right) = \sum_q f_j^q \lambda_j^q. \qquad (5)$$

Note, that individual terms in this sum are unattainable during measurements. Only the whole sum is available for measurement. And that is the reason why we call such measurements fuzzy or mixed.

The methods for the statistical fluctuation analysis of reconstructed states developed in [11, 12, 19, 20] let us perform a detailed description of fidelity, which can be reached for certain quantum measurement protocols. Let us formulate briefly the results for pure quantum states reconstruction. Using the complex vector-state $c$ and complex measurement matrices $\Lambda_j$ one introduces the corresponding entities of double dimension:

$$\tilde{c} = \begin{pmatrix} \text{Re}(c) \\ \text{Im}(c) \end{pmatrix}, \quad \tilde{\Lambda}_j = \begin{pmatrix} \text{Re}(\Lambda_j) & -\text{Im}(\Lambda_j) \\ \text{Im}(\Lambda_j) & \text{Re}(\Lambda_j) \end{pmatrix}.$$

The basic tool for the quantum tomography fidelity analysis is so-called complete information matrix [11, 12, 19, 20]:

$$H = 2\sum_j \frac{t_j}{\lambda_j}\left(\tilde{\Lambda}_j \tilde{c}\right)\left(\tilde{\Lambda}_j \tilde{c}\right)^+.$$

We assume $n$ to be large enough, so we can use the asymptotic theory of statistical estimations. Furthermore, the following normalization condition takes place: $\langle \tilde{c} | H | \tilde{c} \rangle = 2n$.

Let $d\tilde{c}$ be the difference between the exact vector-state and the one reconstructed by the maximum likelihood method (in the real Euclidean space of a doubled dimension). Then the level of statistical fluctuations could be described with the chi-squared distribution [11, 20]:

$$2\langle d\tilde{c} | H | d\tilde{c} \rangle = \chi^2(\nu_H).$$

For the considered case of pure states $\nu_H = 2s - 1$.

Let our measurement protocol form the decomposition of unity and be tomographically complete [20]. Let us also set the normalization condition in the simplest form $\langle c | c \rangle = 1$. Then the eigenvalues $S_{H,j}$ of matrix $H$ allow us to calculate the variances of principal components fluctuations for the reconstructed real vector-state $\tilde{c}$ of double dimension:

$$d_j = \frac{1}{2S_{H,j}}, \quad j = 1, 2, \ldots, \nu, \tag{6}$$

where $\nu$ is the number of quantum state degrees of freedom. For a pure state in the Hilbert space of dimension $s$ the number of degrees of freedom is $\nu = 2s - 2$. Note, that the matrix $H$ is a real symmetric non-negatively definite matrix of dimension $2s \times 2s$. It has $2s$ non-negative eigenvalues. In the case of a tomographically complete protocol only one eigenvalue of the matrix $H$ is equal to zero. It corresponds to the arbitrariness of quantum state global phase and should be rejected. Moreover, one should exclude the maximal eigenvalue, which corresponds to the quantum state normalization condition. The leftover $2s - 2$ eigenvalues of the matrix $H$ define the vector $d$ according to (4).

Vector $d$ is the parameter for the universal fidelity loss distribution which has the form of generalized chi-squared distribution [20]. In this case the loss of reconstruction fidelity $1 - F$ is the following random variable:

$$1 - F = \sum_{j=1}^{\nu} d_j \xi_j^2,$$

where $\xi_j \sim N(0,1)$, $j = 1, \ldots, \nu$ are independent normally distributed random variables with zero means and unit variances.

The fidelity loss average and variance appear to be

$$\langle 1 - F \rangle = \sum_{j=1}^{\nu} d_j, \quad \sigma^2 = 2\sum_{j=1}^{\nu} d_j^2.$$

## 2. DECOHERENCE IMPACT ON MEASUREMENTS

Let us consider an open quantum system, of which the state in the initial time is set by the density matrix $\rho_{in}$. Being open, the system constantly exchanges information with its environment in an uncontrollable way, which results in decoherence of the state under study. This process can be generally described using the operator sum $\mathbf{E}(\rho) = \sum_k E_k \rho E_k^+$ so the state at the output of such channel is defined by the equation $\rho_{out} = \sum_k E_k \rho_{in} E_k^+$ [18, 21]. Here $E_k$ are the elements of transformation, or the Kraus operators.

During the tomography of the state $\rho_{out}$, the probability of the *j*-th event registration, defined by (1), is

$$\lambda_j = \text{Tr}\left(\Lambda_j \rho_{out}\right) = \text{Tr}\left(\Lambda_j \sum_k E_k \rho_{in} E_k^+\right) = \text{Tr}\left(\Lambda_j^{mixed} \rho_{in}\right). \tag{7}$$

Here the mixed measurement operators have been introduced:

$$\Lambda_j^{mixed} = \sum_k E_k^+ \Lambda_j E_k. \tag{8}$$

From the point of view of the events registration probabilities, the tomography of the state $\rho_{out}$ using the protocol with measurement operators $\Lambda_j$ is equivalent to the tomography of the state $\rho_{in}$ using the protocol with measurement operators $\Lambda_j^{mixed}$. This allows us to reconstruct the initial quantum state of the system in the form unaltered by the system-environment interaction. In other words, the decoherence channel becomes the part of the generalized measurement set-up, which operates on the initial state directly (Fig. 1). At the same time, as it will be shown in the next section, the presence of decoherence results in a certain degradation of the reconstruction fidelity. This degradation increases with the increase of the quantum noise level. To compensate this effect one should increase the number of the statistical ensemble representatives. Another condition for the developed method applicability is the need to have the complete information about the noise characteristics in the channel (for example, one needs to know the form of $E_k$ operators or the Choi-Jamilkowski state [18, 21-23]). To get this information in real experiments, one should perform the quantum process tomography for the channel under consideration [23-28].

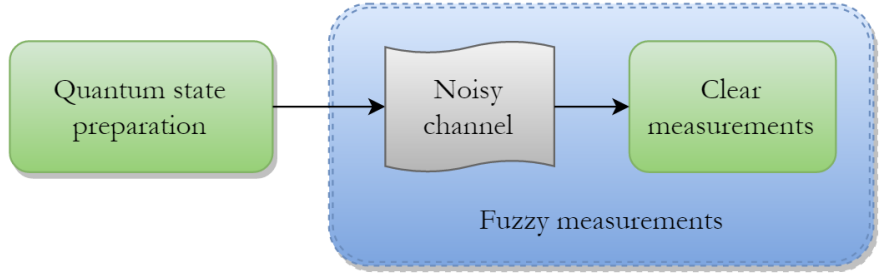

Figure 1. The scheme of the generalized quantum measurement with the quantum noises taken into consideration

It could seem that such a description is rather excessive as one could reconstruct the corrupted state $\rho_{out}$ using the standard approach of clear measurements. After that, one can perform the transform that is inverted to $\{E_k\}$ and get the initial state $\rho_{in}$ (formally this could be done using the evolution matrix [26]). However, this approach has several significant drawbacks. Firstly, channels with decoherence lead to an increase in the rank of a state so one has more parameters to estimate during the reconstruction. Each parameter brings statistical error so the fidelity of the whole state reconstruction will inevitably decrease. Secondly, what is important is that the inverse transform from the output to the initial state is not completely positive in the general case so it will usually lead to negative eigenvalues of the initial state. The reconstruction method presented in this section has no such drawbacks as it directly works with the initial state.

As an example, let us consider the tomography of a qubit, which is subjected to different decoherence mechanisms. It is convenient to represent the *j*-th projection operator with the vector $\mathbf{r}_j = (r_{j,x}, r_{j,y}, r_{j,z})$, $|\mathbf{r}| = 1$, on the Bloch sphere. In this case the *j*-th projection operator for clear measurement is

$$\Lambda_j = \frac{1}{2}\begin{pmatrix} 1+r_{j,z} & r_{j,x}+ir_{j,y} \\ r_{j,x}-ir_{j,y} & 1-r_{j,z} \end{pmatrix}.$$

In the case of the channel with amplitude relaxation having the parameter $T_1$ and acting with duration $t$ the fuzzy measurement operator is as follows:

$$\Lambda_j^{ampl} = \frac{1}{2}\begin{pmatrix} 1+r_{j,z} & (r_{j,x}+ir_{j,y})e^{-t/2T_1} \\ (r_{j,x}-ir_{j,y})e^{-t/2T_1} & 1-r_{j,z}(2e^{-t/T_1}-1) \end{pmatrix}.$$

Similarly, for the process of the pure phase relaxation having the parameter $T_2^{pure}$:

$$\Lambda_j^{phase} = \frac{1}{2}\begin{pmatrix} 1+r_{j,z} & (r_{j,x}+ir_{j,y})e^{-t/T_2^{pure}} \\ (r_{j,x}-ir_{j,y})e^{-t/T_2^{pure}} & 1-r_{j,z} \end{pmatrix}.$$

For the errors of bit-flip and phase-flip that arise with probability $p$, $0 \leq p \leq 1/2$, we have the following operators respectively:

$$\Lambda_j^{bit-flip} = \frac{1}{2}\begin{pmatrix} 1+r_{j,z}(1-2p) & r_{j,x}+ir_{j,y}(1-2p) \\ r_{j,x}-ir_{j,y}(1-2p) & 1-r_{j,z}(1-2p) \end{pmatrix},$$

$$\Lambda_j^{phase-flip} = \frac{1}{2}\begin{pmatrix} 1+r_{j,z} & (r_{j,x}+ir_{j,y})(1-2p) \\ (r_{j,x}-ir_{j,y})(1-2p) & 1-r_{j,z} \end{pmatrix},$$

where $\sigma_x$ and $\sigma_z$ are the Pauli matrices. Note, that the phase-flip error is equivalent to the pure phase relaxation so we will not consider this case further.

In present work for the sake of quantum tomography simulation we use three different quantum measurement protocols, based on the regular polyhedron geometry: tetrahedron, cube and octahedron protocols [19, 29, 30]. This protocols are described with the following instrumental matrices:

$$X^{tetra} = \frac{1}{12^{1/4}}\begin{pmatrix} \sqrt{\sqrt{3}+1} & e^{i\pi/4}\sqrt{\sqrt{3}-1} \\ \sqrt{\sqrt{3}-1} & e^{i3\pi/4}\sqrt{\sqrt{3}+1} \\ \sqrt{\sqrt{3}+1} & e^{i5\pi/4}\sqrt{\sqrt{3}-1} \\ \sqrt{\sqrt{3}-1} & e^{i7\pi/4}\sqrt{\sqrt{3}+1} \end{pmatrix}, \quad X^{cube} = \frac{1}{\sqrt{2}}\begin{pmatrix} \sqrt{2} & 0 \\ 0 & \sqrt{2} \\ 1 & 1 \\ 1 & -1 \\ 1 & i \\ 1 & -i \end{pmatrix}, \quad X^{octa} = \frac{1}{12^{1/4}}\begin{pmatrix} \sqrt{\sqrt{3}+1} & e^{i\pi/4}\sqrt{\sqrt{3}-1} \\ \sqrt{\sqrt{3}+1} & e^{i3\pi/4}\sqrt{\sqrt{3}-1} \\ \sqrt{\sqrt{3}+1} & e^{i5\pi/4}\sqrt{\sqrt{3}-1} \\ \sqrt{\sqrt{3}+1} & e^{i7\pi/4}\sqrt{\sqrt{3}-1} \\ \sqrt{\sqrt{3}-1} & e^{i\pi/4}\sqrt{\sqrt{3}+1} \\ \sqrt{\sqrt{3}-1} & e^{i3\pi/4}\sqrt{\sqrt{3}+1} \\ \sqrt{\sqrt{3}-1} & e^{i5\pi/4}\sqrt{\sqrt{3}+1} \\ \sqrt{\sqrt{3}-1} & e^{i7\pi/4}\sqrt{\sqrt{3}+1} \end{pmatrix}.$$

# 3. EXAMPLES OF NUMERICAL SIMULATION

In Fig. 2 the results of numerical experiment of the pure state reconstruction with the dephasing channel are depicted. Also, the comparison of the numerical experiment and the fidelity loss theoretical estimation, performed by the use of the complete information matrix, briefly described in the previous section, is presented. It is seen from the pictures that increasing the decoherence duration leads to the shift of fidelity loss distribution to the higher values.

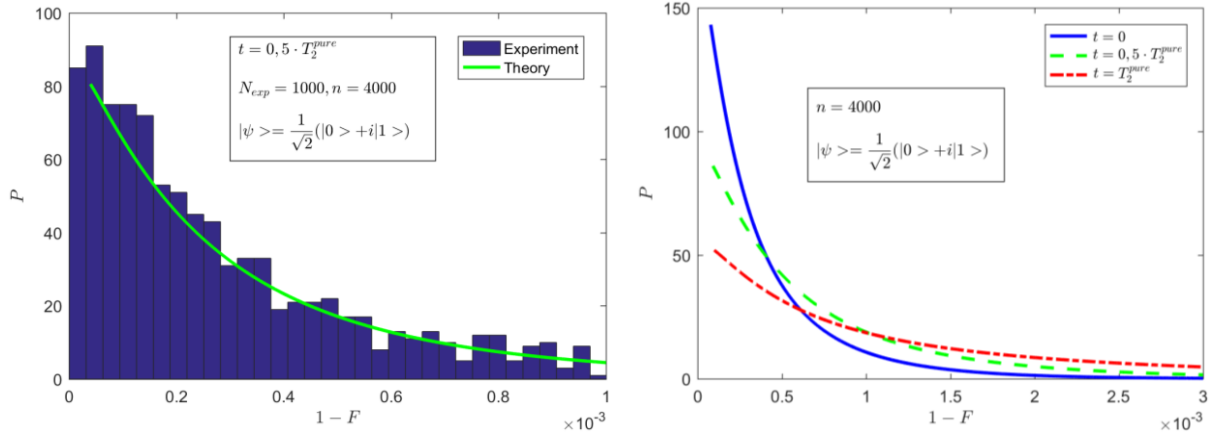

Figure 2. Left – the result of numerical simulation of the pure state $|\psi\rangle = (|0\rangle + i|1\rangle)/\sqrt{2}$ reconstruction using the tetrahedron protocol with the impact of the pure phase relaxation of duration $t = 0,5 \cdot T_2^{pure}$; green line depicts the theoretical estimation of fidelity loss distribution, which is made by using the complete information matrix; the number of numerical experiments is 1000; the sample size in each experiment is 4000. Right – theoretical fidelity loss distribution for different decoherence durations.

To estimate the states reconstruction fidelity loss it is convinient to use the following measure:

$$L = n\langle 1-F \rangle = n\sum_{j=1}^{\nu} d_j,$$

which does not depend on the sample size $n$ as the vector $d$ is inversely proportional to $n$.

In Fig. 3 the distribution of this quantity on the Bloch sphere for the tetrahedron, cube and octahedron protocols, noised by the amplitude relaxation, pure phase relaxation, bit-flip and phase-flip errors, is depicted (each point on the sphere corresponds to the pure qubit state; the color of this point marks the magnitude of $L$ for the given protocol). In the case of clear measurements minimum $L_{min}$ and maximum $L_{max}$ values of $L$ are invariant to the protocol orientation. As it is seen from the figure the presence of decoherence leads to the anisotropy of the average fidelity loss distribution on the Bloch sphere. As a result, the protocol orientation influences the distribution of $L$ and its magnitude. For example, for the cube protocol that is used in this work in the case of the pure phase relaxation with duration $t = 0,8 \cdot T_2^{pure}$: $L_{min} \approx 4,09$ and $L_{max} \approx 5,57$. Rotation of the protocol around the axis $\mathbf{n} = (1/\sqrt{2}, 1/\sqrt{2}, 0)$ by the angle $\pi/4$ leads to the reduction of fidelity losses average: $L_{min} \approx 3,7$ and $L_{max} \approx 5,13$. Now, the amplitude relaxation with the duration $t = 0,8 \cdot T_2^{pure}$ results the values $L_{min} \approx 5,04$ and $L_{max} \approx 9,92$. If we perform the same rotation, as in the previous case, we will get $L_{min} \approx 4,51$ and $L_{max} \approx 14,61$.

The described analysis of the quantum noise impact on reconstruction fidelity lets us estimate, how much one should increase the sample size to maintain the reconstruction fidelity in the presence of decoherence. For example, for the clear tetrahedron protocol the maximum average fidelity loss is $L_{max}^{ideal} = 1,5$. Yet, having the amplitude relaxation of duration $t = 1,5 \cdot T_1$, this value increases to $L_{max}^{mixed} \approx 17,25$. To compensate this effect, one should increase the sample size by $L_{max}^{mixed} / L_{max}^{ideal} \approx 11,5$ times.

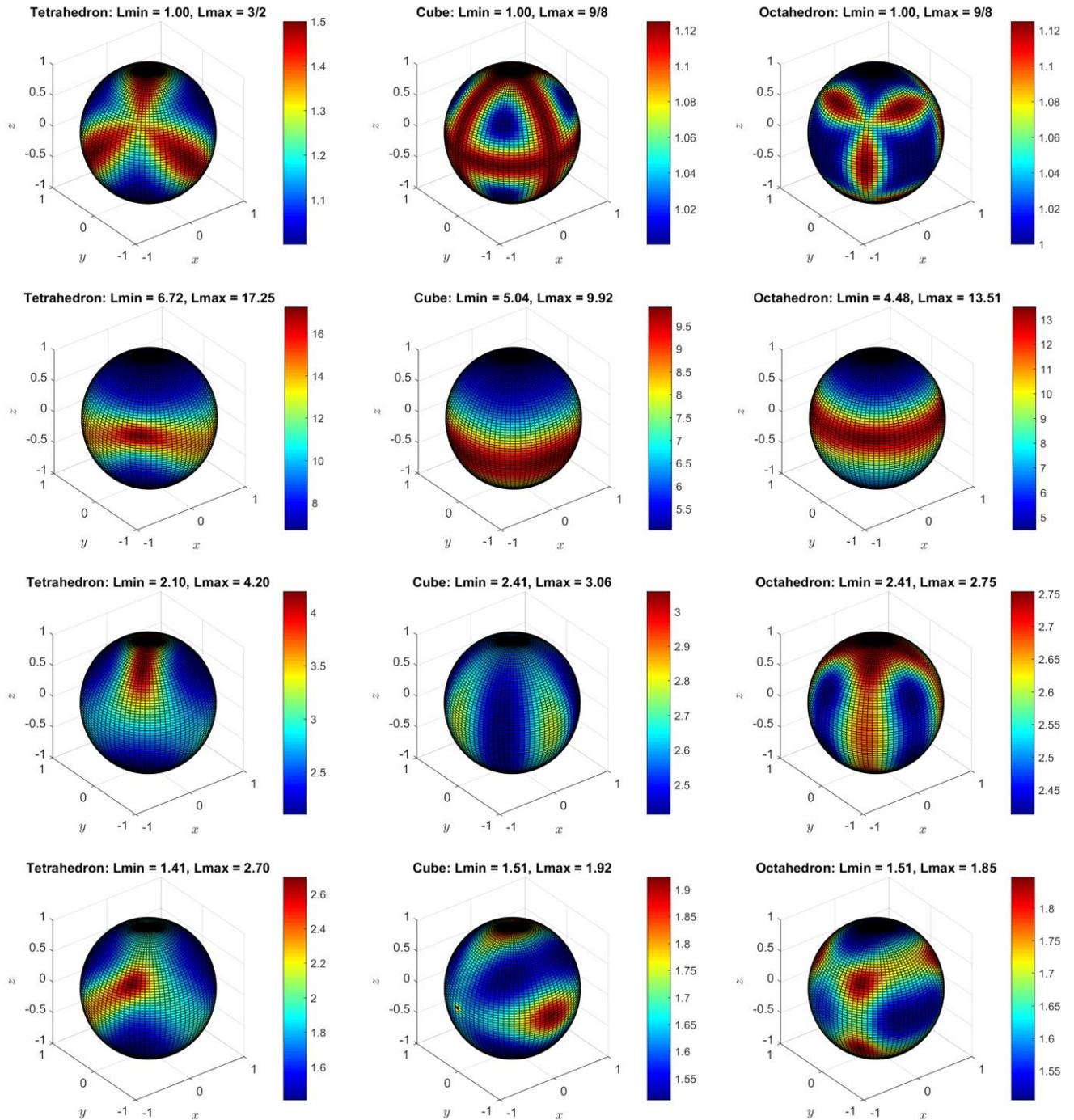

Figure 3. The average fidelity loss distribution on Bloch sphere for tetrahedron (left), cube (middle) and octahedron (right) protocols. From the top: ideal, amplitude relaxation ($t = 1,5 \cdot T_1$), phase relaxation ($t = 0,5 \cdot T_2^{pure}$) and bit-flip error ($p = 0,1$) cases.

For the sake of tomography of *n*-qubit quantum states we shall introduce the instrumental matrix derived as *n* tensor products of single qubit instrumental matrices (thereby we consider the projection on the non-entangled states only):

$$X^{\otimes n} = \underbrace{X \otimes X \otimes \ldots \otimes X}_{n}.$$

For simplicity, we also consider the non-entangling quantum channels with decoherence, where on the *j*-th qubit, independent from of the other qubits' states, there acts a process, defined by the set of Kraus operators $\{E_{k_j}\}$. Then the mixed measurement operators are as follows:

$$\Lambda_j^{mixed} = \sum_{k_1, k_2, \ldots, k_n} \left(E_{k_1}^+ \otimes \ldots \otimes E_{k_n}^+\right) \Lambda_j \left(E_{k_1} \otimes \ldots \otimes E_{k_n}\right).$$

Let us consider for example the reconstruction of a 2-qubit state. In Fig. 4 the results of numerical experiment of the pure entangled state reconstruction using the 2-qubit tetrahedron protocol are depicted. Here the protocol is noised with the amplitude relaxation acting on the first and second qubits with the durations $t^1$ и $t^2$ respectively. The time of the amplitude relaxation is $T_1$ for both qubits.

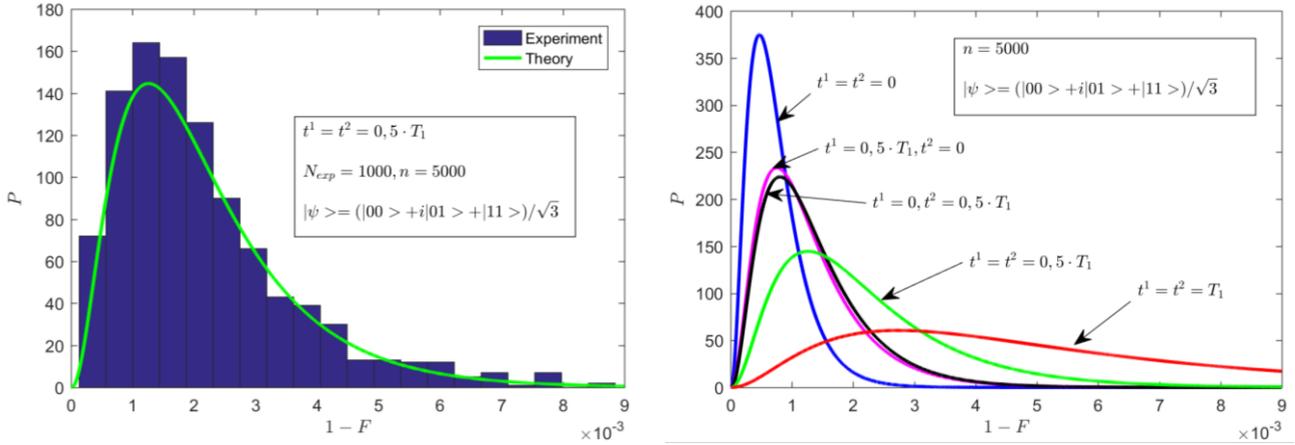

Figure 4. Left – the result of numerical experiment of the pure entangled state $|\psi\rangle = (|00\rangle + i|01\rangle + |11\rangle)/\sqrt{3}$ reconstruction using the 2-qubit tetrahedron protocol noised with the amplitude relaxation of both qubits with duration $t^1 = t^2 = 0{,}5 \cdot T_1$; green line – theoretical estimation of fidelity loss distribution; the number of experiments is 1000; the sample size in each experiment is 5000. Right – theoretical fidelity loss distribution for different qubits decoherence duration.

## 4. CONCLUSION

In the present work the formalism of mixed (fuzzy) measurements, which is the generalization of the pure projective measurements, has been considered. This formalism lets us perform the quantum states tomography, when the measurements of different projections are realized with different probabilities, but we have access to the aggregate statistics only. The important case of the fuzzy measurements origination in the presence of the quantum state decoherence, has been studied. Herein the decoherence channel is considered to be the part of the measurement set-up by modifying the measurement operators. Such approach lets us reconstruct with high fidelity the input state in the uncorrupted form, before it started to decohere. In particular, we are able to reconstruct a pure state of the system, even if it had been highly damaged before it started to interact with the detector. However, the presence of decoherence results in a decrease in the reconstruction fidelity, so one has to increase the sample size to compensate this effect. It has been also noted that the protocol orientation on the Bloch sphere influence the average fidelity losses of the quantum states reconstruction. The research presented in this work is important for the control of fidelity of real quantum information technologies where decoherence is one of the main problems on the way of its practical realization.

This work was supported by the Program of the Russian Academy of Sciences in fundamental research.